# Improving Performance of Tin-Doped-Zinc-Oxide Thin-Film Transistors by Optimized Multi-Stacked Active-Layer Structures

Zhuofa Chen†, Dedong Han†*, Xing Zhang† and Yi Wang†

†Institute of Microelectronics, Peking University, Beijing 100871, China

* Corresponding author.

*E-mail: handedong@pku.edu.cn

## Abstract

In this paper, we investigated the performance of thin-film transistors (TFTs) with different channel configurations including single-active-layer (SAL) Sn-Zn-O (TZO), dual-active-layers (DAL) In-Sn-O (ITO)/TZO, and triple-active-layers (TAL) TZO/ITO/TZO. The TAL TFTs were found to combine the advantages of SAL TFTs (a low off-state current) and DAL TFTs (a high mobility and a low threshold voltage). The proposed TAL TFTs exhibit superior electrical performance, e.g. a high on-off state current ratio of $2 \times 10^8$, a low threshold voltage of 0.52 V, a high saturation mobility of 145.2 cm$^2$/Vs, and a low off-state current of 3.3 pA. The surface morphology and characteristics of the ITO and TZO films were investigated and the TZO film was found to be C-axis-aligned crystalline (CAAC). A simplified resistance model was deduced to explain the high channel resistance of TAL TFTs. At last, TAL TFTs with different channel lengths were also discussed to show the stability and the uniformity of our fabrication process. Owing to its low-processing temperature, superior electrical performance, and low cost, TFTs with the proposed TAL channel configuration are highly promising for flexible displays where the use of heat-sensitive polymeric substrates is desirable.

## Introduction

Thin-film transistors (TFTs) has been widely used as switching devices for flat-panel display such as active-matrix liquid crystal displays (AM-LCD) and Active Matrix Organic Light Emitting Diodes (AM-OLED). High-performance TFTs with a high mobility, a low threshold voltage, and a low swing slope can reduce the power consumption and enhance the quality of flat-panel display.[1-4] Therefore, various studies has be carried out to improve the electrical performance of TFTs, this includes adopting different device structures,[5] using different materials for the channel layer, and optimizing the fabrication processes.[6]

TFTs fabricated by solution processing and inkjet printing have the advantage of low cost, while suffering from a low mobility and a high annealing temperature.[7,8] TFTs based on 2-dimentional (2D) materials such as graphene and Molybdenum disulfide ($MoS_2$) have been widely investigated recently due to their excellent electrical properties.[9,10] However, 2D materials-based TFTs still have some challenges in large-scale fabrication of high quality devices, not compatible with modern Silicon-based microelectronic technologies. Zinc-oxide (ZnO) based TFTs have attracted considerable attention for their superior electrical and optical properties since last decade.[2,3,11-13] Among ZnO-based multicomponent oxide TFTs, In-Ga-Zn-O, Al-Zn-O, In-Zn-O, Zn-In-Sn-O TFTs had been proved to be attractive alternatives to conventional silicon-based TFTs in AMOLED due to their high mobility, low threshold voltage, fully transparency, and large-area applications.[14-20] While most of these work required a high processing or annealing temperature (above 300 ℃). These thermal processes increase the manufacturing cost and limits their application in flexible display where low processing temperature (<100 ℃) is desirable.[14,21,22] Thus, alternative ZnO-based TFTs fabricated at low temperature still need to be investigated. Sn-doped ZnO (TZO) has the advantages of high mobility and low temperature processing compatibility.[23,24] While the research of TZO TFTs received less attention and the device performance presented is undesirable. High-performance TZO TFTs fabricated at a low temperature are still of interest. Therefore, the goal of our research is to realize high-performance TZO TFTs at a low temperature.

Multi-stacked active-layer structures have been previous demonstrated to improve the performance of solution-processed TFTs.[25,26] However, a systematic work to probe the performance of TZO TFTs with multi-layer is still lacking. Previously, we reported improving the performance of TZO TFTs with various strategies such as adding oxygen during the deposition of TZO layers,[27,28] adopting DAL ITO/TZO TFTs,[29,30] and adjusting the thickness of the ITO/TZO active layer.[31] We demonstrated that TZO TFTs are promising switching devices for flat-panel display. The DAL TFTs can effectively improve the mobility and reduce the threshold voltage. However, the DAL TFTs have a high off-state current due to the high carrier density in the ITO layer, leading to a higher power consumption in real applications. Therefore, we aim to optimize the channel configuration of the TFTs to reduce the off-state current and improve the on-off current ratio.

In this paper, we compared the performance of TFTs with different channel configurations and demonstrated that high-performance TZO TFTs can be realized at low temperature

(80 °C) by adopting TAL stack for TFTs. Compared to TFTs with SAL or DAL channel configuration, the proposed TAL TZO/ITO/TZO TFTs exhibit a saturation mobility ($\mu_{sat}$) of 145.2 cm$^2$/Vs and a low threshold voltage ($V_{th}$) of 0.52V. The quality of the TZO film and ITO films were characterized by AFM, SEM, and XRD. The stability and uniformity of our fabrication process is confirmed by the consistent performance of TAL TFTs with different channel lengths. A physical mechanism for the electrical improvement is also deduced. The proposed TAL TFTs are promising in various applications due to the superior performance, low-processing temperature, and low cost.

## Results

**Device structure and fabrication process.** A schematic of the device structure is shown in Fig. 1a. A bottom-gate TFT was fabricated on glass substrate by standard photolithography and lift-off techniques, without any intentional substrate-heating. All procedures were carried out with a temperature below 80°C. A top-view optical image of a representative device is shown in Fig. 1b. The device was fabricated using a 3 photo-masks process, as shown in Fig. 2. The fabrication procedures are described in methods.

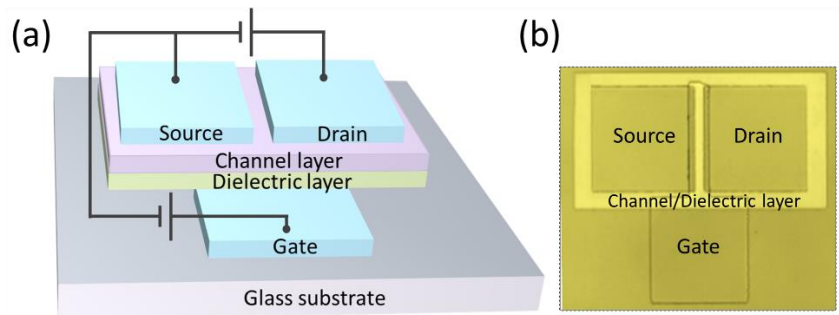

**Figure 1.** (a) Schematic illustration of the device structure. An inverted staggered structure was adopted in this research. The channel layer and the dielectric layer are patterned using the same mask. (b) An optical photo (top view) of a representative device in this paper.

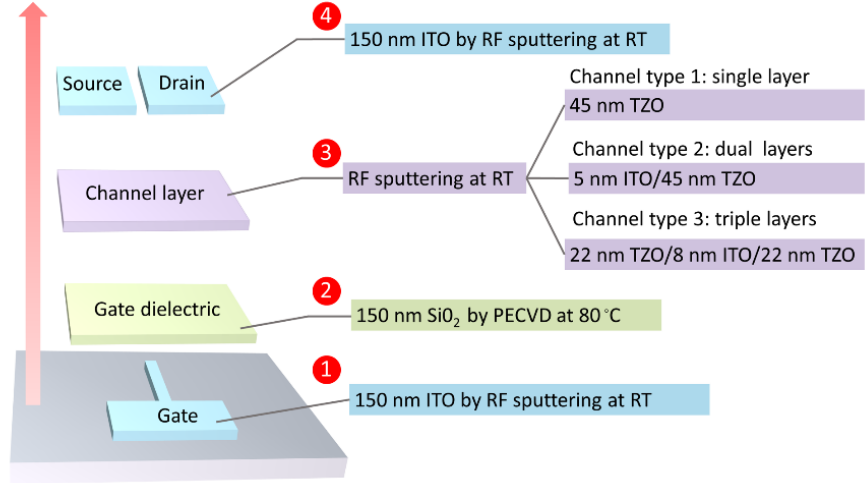

**Figure 2.** Fabrication process of the TFTs with three different channel configurations: channel type 1 (SAL), channel type 2 (DAL), and channel type 3 (TAL). The devices were fabricated from step 1 to step 4 successively. Three different kinds of devices were fabricated with the corresponding channel type (SAL, DAL, and TAL).

**Electrical measurements.** Fig. 3a-c shows schematics of three different channel configurations: SAL, DAL, and TAL. Fig. 3d shows the representative transfer curves of TFTs with three different channel configurations: TZO/ITO/TZO (TAL), ITO/TZO (DAL), and TZO (SAL). All the devices have the same channel dimension with a channel width of 100 $\mu m$ and a channel length of 20 $\mu m$. The drain to source voltage was biased at 5V to make sure the TFTs operate on saturation region. Back-gate voltage was biased from -4V to 10 V. The transport measurements were carried out under ambient condition at room temperature. Fig. 3d shows that the TAL TFTs have the best performance with a high on-off state current ratio ($I_{on}/I_{off}$) of ~ $2 \times 10^8$ and a low $V_{th}$ of ~0.5 V. Moreover, TAL TFTs has a high $\mu_{sat}$ of 145.2 cm$^2$/Vs and a low $I_{off}$ of 3.3 pA. The Subthreshold Slope (*SS*) was calculated by the equation (1), while $V_{th}$ and $\mu_{sat}$ were extracted by the equation (2) and (3), using the linear fitting based on 10% - 90% of the maximum $I_{DS}$ on the $I_{DS}^{1/2}$ versus $V_{GS}$ plot [32]. $C_{ox}$ of 2.6×10⁻⁸ F/cm$^2$ was extracted from C-V curve of 100K Hz[33].

$$SS = \frac{\partial V_{GS}}{\partial (log I_{DS})}/V_{DS}=con \qquad (1)$$

$$I_{DS}^{\frac{1}{2}} = \left(\frac{W}{2L}\mu_S C_{ox}\right)^{\frac{1}{2}}(V_G - V_{th}) \qquad (2)$$

$$\mu_s = \frac{2L}{WC_{ox}}\left(\frac{\partial I_{DS}^{\frac{1}{2}}}{\partial V_G}\right)^2 \qquad (3)$$

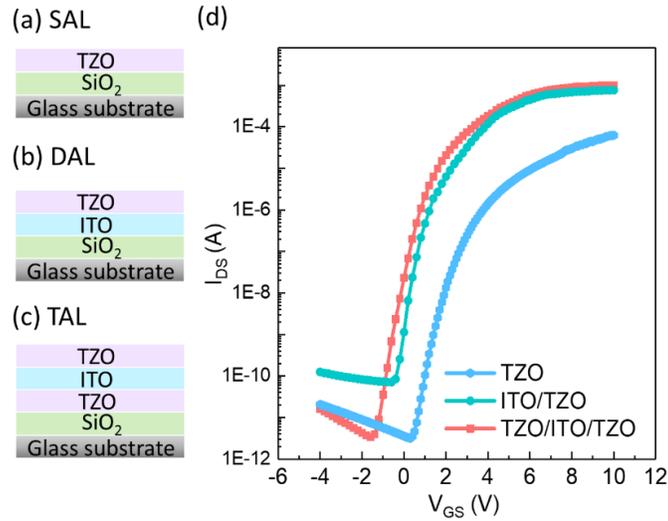

**Figure 3.** (a) Schematic of TZO single-active layer, (b) Schematic of ITO/TZO dual-active layers, (c) Schematic of TZO/ITO/TZO triple-active layers. (d) Representative transfer curves of TFTs with the three different channel configurations: SAL, DAL, and TAL. The drain to source voltage was set to 5V.

Fig. 4 compares the electrical properties of TFTs with three different channel configurations. Fig. 4a compares the $μ_{sat}$ of the devices. We can see that comparing to SAL TZO TFTs, the TFTs with TAL and DAL channel configurations have much higher $μ_{sat}$ (roughly 5 times higher). This high mobility is due to the good conductivity of the ITO layer in the channel[34]. Fig. 4b compares the $I_{on}/I_{off}$ and $V_{th}$ and shows that the TAL TFTs have the lowest $V_{th}$ and highest $I_{on}/I_{off}$. The DAL TFTs and SAL TFTs has similar $I_{on}/I_{off}$. After adding the ITO layer, both the $I_{on}$ and $I_{off}$ of the DAL TFTs are increased. Compared to SAL TFTs, DAL TFTs has the advantages of high $μ_{sat}$ and low $V_{th}$ while also suffering from the disadvantage of high $I_{off}$. The TAL TFTs combine the advantages of SAL TFTs (low $I_{off}$) and DAL TFTs (high $μ_{sat}$ and low $V_{th}$). Fig. 4c shows the variation of *SS* due to back gate voltage in TFTs with TAL, DAL, and SAL, respectively. All the TFTs have similar values of *SS* (~0.3V/dec.). Fig. 4d shows the channel resistivity of the SAL stack, DAL stack, and TAL stack. The SAL stack and TAL stack have roughly the same channel resistivity, ~20 times larger than that in DAL stack. This confirms the lower $I_{off}$ in TAL and SAL TFTs while the higher $I_{on}$ in DAL TFTs, shown in Fig. 3d. Extracted parameters were summarized in Table 1.

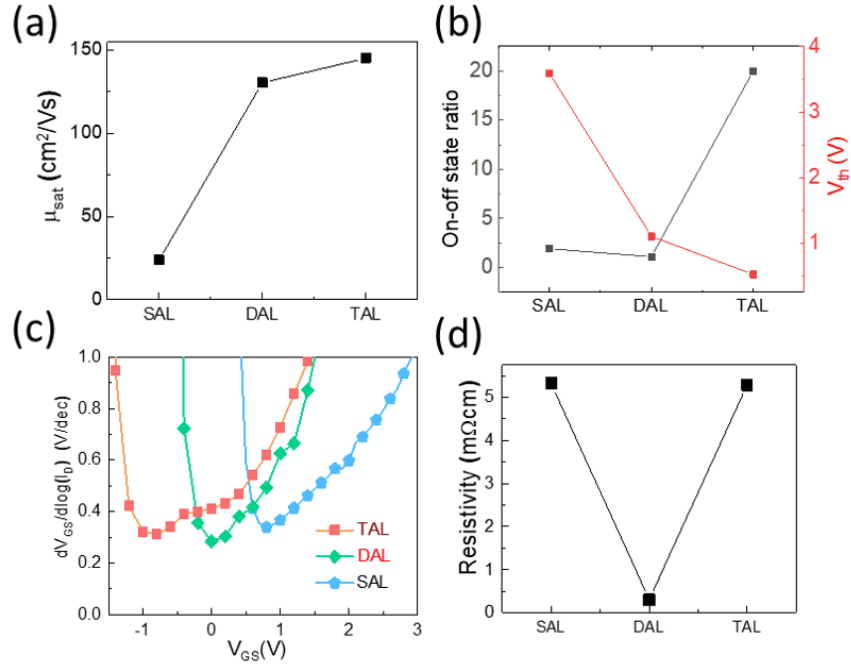

**Figure 4.** Comparison of the electrical properties in three different TFTs. (a) Saturation mobility, (b) on-off state current ratio and threshold voltage, (c) subthreshold slope, and (d) channel resistivity.

These electrical results are originated from the different roles of each film in the channel. As the n-channel TZO TFTs operated on enhancement mode, most of the induced carriers go either into the deep localized states in the TZO layer or into the interface states when the gate bias voltage $V_{GS}$ < 0V. Only a very small fraction of electrons that are close to the front of TZO/SiO$_2$ interface (interface near to the gate electrode) participate in channel conduction, resulting in a low $I_{off}$. While as the $V_{GS}$ increases, the channel conductivity increases rapidly due to charges accumulating in the TZO layer, yielding a suitable high $I_{on}$. The TZO channel controls the charge conductance to get a high $I_{on}/I_{off}$ and a suitable $V_{th}$. While for DAL ITO/TZO TFTs, the ITO layer dominates in $I_{on}$ transmission for its high carrier density, leading to a higher $I_{on}$. The TZO layer provides a suitable $V_{th}$ and $I_{on}/I_{off}$ ratio for its low carrier density on off-state and its controlling ability in the charge conductance. Compared to TZO conducting layer, the thin ITO layer of the DAL ITO/TZO channel provides a higher carrier concentration, therefore maximizing the charge accumulation and yielding a high $μ_{sat}$.[35]

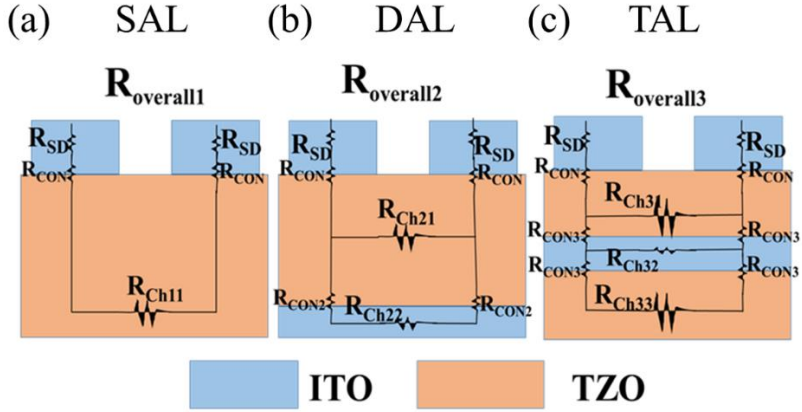

**Figure 5.** Schematic cross-sectional view of the overall resistance in (a) SAL TZO TFTs, (b) DAL ITO/TZO TFTs and (c) TAL TZO/ITO/TZO TFTs. $R_{SD}$ is the resistance in the source and drain electrode, $R_{CON}$, $R_{CON2}$, $R_{CON3}$ are the interface resistances between interface and $R_{Ch11}$, $R_{Ch21}$, $R_{Ch22}$, $R_{Ch31}$, $R_{Ch32}$, $R_{Ch33}$ are the resistances in the active layers.

From the resistance point of view, the ITO layer reduces the channel resistance of ITO/TZO TFTs while encapsulating the ITO layer between two TZO layers can increases the channel resistance. The schematic illustration of the three different channel configurations is shown in Fig. 5. Compared to the SAL TZO TFTs, the higher carrier density in the ITO layer, leads to smaller channel layer resistance $R_{ch22}$ and $R_{ch32}$ (shown in Fig. 5b-c)[36], resulting in smaller overall resistance of ITO/TZO TFTs ($R_{overall2}$) despite small contact resistance $R_{con}$ and $R_{con2}$. Using the equation (4):

$$I_{off} = \frac{V_{DS}}{R_{overall}} \qquad (4)$$

While for the TAL, thinner TZO layer has lower carrier density, yielding larger channel resistances $R_{ch31}$ and $R_{ch33}$, the series resistance $R_{con3}$ also adds to $R_{overall3}$ (shown in Fig. 5(c)).[31,37] Thus, $R_{overall3}$ is larger than the $R_{overall2}$. This can be confirmed by the resistivity shown in Fig. 4d. Therefore, the TAL TFTs have lower $I_{off}$ than the DAL TFTs.

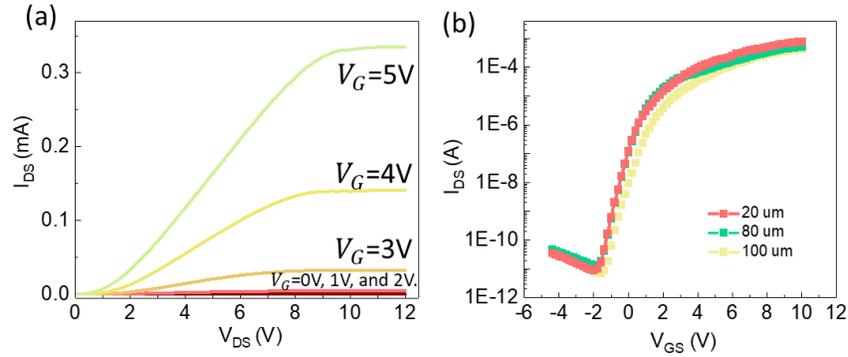

Figure 6. (a) . Output characteristics of the TAL TFTs. (b) Representative transfer characteristics of TAL TFTs with channel length of 20 μm, 80 μm, and 100 μm.

**Output characteristics and device stability.** Fig. 6a shows the output characteristics of the TAL TFTs. The TAL TFTs work on enhancement mode and the back-gate voltage was set from 0V to 5 V with a step of 1 V. The drain and source voltage scans from 0V to 12V. The drain current is raised rapidly within 1 V between drain and source and a clear saturation region can be observed. This demonstrates the good switch controlling ability (switch from off-state to on-state rapidly) of the device. Fig. 6a shows that the saturation current exceeds 300 μA at a low back-gate voltage of 5 V. This indicates good current driving ability in the TAL TFTs. However, nonlinear correlation between the $V_{DS}$ and the $I_{DS}$ was also observed for $V_{DS}$ <1V. This may be due to the parasitic resistance induced by trap states near source and drain regions, leading to the current crowding phenomenon. Part of the drain voltage may drop on the parasitic resistance.[38] Due to the limitation of our setup, all the electrical characteristic measurements were performed under ambient condition. Oxygen may be adsorbed on the top of the channel and form a depletion layer. This may also lead to current crowding phenomenon. More work can be done to improve the quality of the contact interface but that`s out of the scope of this paper. Moreover, later work will optimize the device structure by adding an insulating layer on top of the channel to prevent this problem.

To investigate the stability and the uniformity of our fabrication process. TAL TFTs with different channel lengths are also fabricated and measured. Fig. 6b shows the representative transfer characteristics of TAL TFTs with three different channel length 20 μm, 80 μm, and 100 μm. The related parameters were extracted and shown in Table 1. All the devices have comparable saturation mobility higher than 100 cm$^2$/Vs and high on-off state current ratio higher than $10^8$. This indicates our fabrication process is stable and uniform.

Table 1 Extracted parameters of the studied TFTs.

| Channel stacks | Chanel width/Length | Thickness (nm) | $\mu_{sat}$ (cm$^2$/V·s) | SS mV/dec. | $V_{th}$ (V) | On-off Ratio |
|---|---|---|---|---|---|---|
| TZO | 100/20 | 45 | 23.8 | 341 | 3.59 | 1.9×10$^7$ |
| ITO/TZO | 100/20 | 5/45 | 130.3 | 282 | 1.1 | 1.1×10$^7$ |
| TZO/ITO/TZO | 100/20 | 22/8/22 | 145.2 | 312 | 0.52 | 2×10$^8$ |
| TZO/ITO/TZO | 100/80 | 22/8/22 | 135.7 | 353 | 0.87 | 1.0×10$^8$ |
| TZO/ITO/TZO | 100/100 | 22/8/22 | 127.9 | 380 | 0.92 | 1..1×10$^8$ |

**Material surface morphology and transparency.** Fig. 7a-b show the AFM surface morphology of the ITO and TZO film, respectively. The RMS is 0.8nm and 1.9nm, respectively. The smooth surface of the ITO film indicates better conductance of the film while the TZO film has a granular surface morphology with a larger surface roughness. The X-ray diffraction in Fig. 7c has one prominent peak at 34.3°, indicating Sn atoms can successfully replace Zn sites in the lattice and form C-axis-aligned crystalline (CAAC).[39-41] The average grain size of the TZO film is estimated to be 17.1 nm using the Scherer formula, this can also be confirmed by the SEM image shown in Fig. 7d.

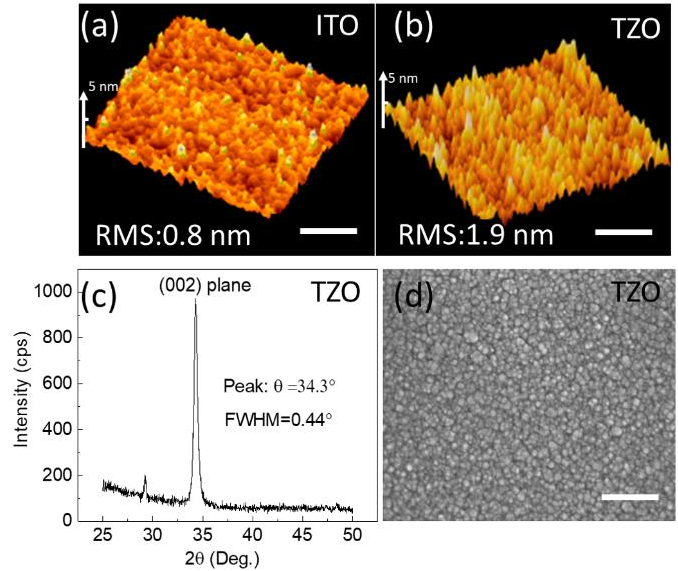

**Figure 7.** Surface characterization off the ITO and TZO films. (a) AFM image of the ITO film. (b) AFM image of the TZO film. (c) XRD diffraction pattern of the TZO film. (d) SEM micrograph of the TZO film. The scale bar for AFM images and SEM images is 200 μm.

## Discussion

For SAL TZO TFTs, oxygen was intentionally added during the RF sputtering process of the TZO film to reduce oxygen vacancy in the material, leading to reduction the hole density in the channel, which can reduce the off-state current and improve the swing slope of the device.[28] This can explain the low off-state current in SAL TZO TFTs. For the DAL ITO/TZO TFTs, ITO layer with a high carrier density was introduced to form channel layer. The high carrier density improves the mobility and the on-state current. Though the DAL ITO/TZO TFTs have superior performance including a high mobility, a low $V_{th}$, and a low *SS*, the high off-state current will lead to a high power consumption in real applications. The TAL channel configuration proposed in this paper has lower off-state current and still maintain a high mobility, can effectively solve this problem. Note that the thickness of the channel stack can also affect the performance of the devices. We have previously reported TZO TFTs and ITO/TZO TFTs with various TZO film thickness and ITO film thickness.[30,31] The thickness of the channel layers of the SAL TFTs and DAL TFTs in this research has been optimized. Thus, we can eliminate the effect of channel thickness when comparing the performance of devices with three different channel configurations. A more systematic work on optimizing the thickness of TAL stacks is on going. But this would not affect our comparison of the three channel configurations and demonstration of the superior performance of the TAL TFTs.

## Conclusions

In this paper, we compared the electrical properties of TFTs with three different channel configurations including SAL, DAL, and TAL. Compared to SAL TFTs, DAL TFTs has a higher mobility and a lower *SS* due to the high carrier density from the ITO layer. While DAL TFTs suffer from a high off-state current, which leads to a higher power consumption in real application. The TAL TFTs were proposed to solve this problem. The proposed TAL TFTs combine the advantages of both SAL TFTs and DAL TFTs and exhibit superior electrical performance such as a high on-off state current ratio of $2 \times 10^8$, a low $V_{th}$ of 0.52 V, a high $\mu_{sat}$ of 145.2 cm$^2$/Vs, and a low off-state current of 3.3 pA. The surface morphology of the ITO and TZO film are investigated. A simplified resistance model was deduced to explain the high resistivity of the TAL channel. A more systematic work on optimizing the thickness of each layer in the TAL stacks is on going. Owing to its advantages of low-processing temperature, superior electrical performance, simple process and low cost, TFTs with the proposed TAL channel configuration are highly promising for oxide semiconductor TFTs manufacturing and have application in flexible displays where the use of heat-sensitive polymeric substrates is desirable. Thus, this investigation is very crucial for commercial applications.

## Methods

**Device fabrication.** The fabrication procedures are described as follows: (1) A gate electrode was patterned and a 150-nm thick ITO film was deposited by radio frequency (RF) magnetron sputtering at room temperature (RT) in Ar (pressure: 1.2 Pa and power:

70W). (2) A 150-nm thick $SiO_2$ was grown using plasma-enhanced chemical vapor deposition (PECVD) with a mixture of $SiH_4$ and $N_2O$ (ratio 65:130) at 80 ℃. (3) Channel layers were deposited by RF sputtering at room temperature in $Ar/O_2$ mixture (flow rate ratio 100/8) with a power of 70W. The target adopted for sputtering was a ceramic target with a mass ratio of $ZnO : SnO_2 = 97 : 3$. In this paper, TFTs with three different channel configurations were fabricated. (a) Single-active-layer TFTs (SAL TFTs) with single TZO layer (channel type 1 in Fig. 2), a 45-nm thick TZO was growth by RF sputtering. (b) Dual-active-layer TFTs (DAL TFTs) with ITO/TZO stack (channel type 2 in Fig. 2), a 5-nm thick ITO was first deposited and followed by depositing a 45-nm thick TZO. (c) Triple-active-layer TFTs (TAL TFTs) with TZO/ITO/TZO stack (channel type 3 in Fig. 2), 22-nm thick TZO, 5-nm ITO, and 22-nm TZO were deposited sequentially by RF sputtering. (4) After patterning the source and drain electrodes, a 150-nm thick ITO film was RF sputtered and lifted to form the source and drain electrodes.

**Device Measurement and Materials Characterizations.** The surface morphology of the TZO films and ITOs films were evaluated by atomic force microscopy (AFM) and scanning electron microscope (SEM). The structure of the TZO film was analyzed by X-ray powder diffraction (XRD). The channel resistivity was obtained from four-probe station. The transport properties of the TFTs were characterized by a semiconductor parameter analyzer (Agilent 4156C).

**Data Availability**. The datasets generated during and/or analyzed during the current study are available from the corresponding author on reasonable request

## Acknowledgements

This work was supported partly by the National Basic Research Program of China (973 program, Grant No. 2013CBA01604) and partly by the National Natural Science Foundation of China (Grant No. 61275025).